\documentclass[superscriptaddress,nofootinbib,twocolumn]{revtex4}
\usepackage{amsfonts} 
\usepackage{graphicx} 
\usepackage{amsmath}  
\usepackage[english]{babel}
\usepackage[latin1]{inputenc}
\usepackage{latexsym}
\usepackage{color}

\begin{document}

\author{E. López Fune}
\email{elpezfun@lpnhe.in2p3.fr}
\affiliation{Laboratoire de Physique Nucl\'{e}aire et de Hautes \'{E}nergies (LPNHE), \\
              4 Place Jussieu, Tour 22, 1er \'{e}tage, 75005 Paris, France.}

\title{Magnetized strange quark matter under stellar equilibrium and finite temperature}

\begin{abstract}
The strange quark matter under strong magnetic fields and finite temperatures is studied in the framework of the MIT Bag model. Matter under such conditions is believed to be present in the core of dense astrophysical objects, like Neutron Stars and more exotic compact objects like Quark Stars. In this study, the anisotropy of the pressure due to the presence of a strong magnetic field is taken into account and a temperature-dependent equation of state is obtained. In the strong field regime, the behavior of the transversal pressure suggests a transversal collapse of the quark and electron gasses for magnetic fields above $\sim10^{19}$ G, even at finite temperature, which can enhance as well the collapse. The corresponding behavior of the energy per baryon and the mass-radius relation for Quark Stars at different temperatures, fixed magnetic field and taking into account the baryon number conservation, $\beta$-equilibrium and charge neutrality, are as well reported.
\end{abstract}

\keywords{strange quark matter, strong magnetic fields, finite temperature.}

\maketitle

\section{Introduction}

The quark matter, as a stable form of matter, has been a subject of intense studies since the construction of the quark model of hadrons by M. Gell-Mann \cite{GellMann:1964nj}. It is considered as a unconfined plasma state made of quarks and gluons, which results as a direct consequence of the asymptotic freedom of QCD, when the baryonic density or temperature are high enough to consider quarks and gluons more fundamental than hadrons. Depending on the temperature and/or the chemical potential of the plasma, this form of matter could appear in daily Nature essentially in two regimes. One could be the Quark Gluon Plasma (QGP), which would be a hot phase of quark matter when $T\gg\mu.$ Our Universe at the beginning, surely would have passed through this phase a few seconds after the Big Bang, when the temperatures were extremely high and the baryon density very low. The QGP could also be created artificially in colliders of heavy ions; in this case, both, the temperature and density of the plasma, depend on the energy of the colliding particles. The other regime is at low temperatures and high density environments $T\ll\mu$ like in the core of heavy Compact Objects (CO), such as Neutron Stars (NS) and the more exotic Quark Stars (QS).

In the core of NSs, at very high densities and low temperatures, neutron matter is believed to experience a phase transition from a neutron liquid to a gas of unconfined quarks and gluons. The $u,\,d-$quarks, along with electrons and neutrinos would produce $s-$quarks through weak interaction processes with a dynamical chemical equilibrium among the constituents, at time scales compared to the life of the Universe itself. This new form of matter is known in the literature as Strange Quark Matter (hereafter SQM) and the NSs will become Strange Quark Stars (SQS) \cite{Weber}. 

At very high densities, zero temperature and zero pressure, conditions that might be perfectly achieved in the interior of NSs \cite{Ivanenko:1965dg, Ivanenko:1969gs, Itoh:1970uw}, the SQM is conjectured to be the ground state of nuclear matter, since by Pauli exclusion principle, it is energetically favourable to produce a new heavy flavour by gaining stability \cite{Bodmer:1971we, Witten:1984rs, Terazawa}. The confirmation of the existence of the SQM could establish a direct connection between the hypothetical SQSs, with the explanations of recent observations of Soft Gamma-ray Repeaters and anomalous radiations from X-pulsars \cite{Weber}.

On the other hand, there are no doubts about the importance of strong magnetic fields in the astrophysical environment. The emission of intense sources of X-rays associated to Pulsars, Magnetars and NSs in general, suggests surface magnetic fields around $10^{13}-10^{15}$ G \cite{Duncan:1992hi, Kouveliotou:1998ze}. In the interior of these COs, the equipartition theorem predicts magnetic fields even higher, of the order of $10^{19}\sim10^{20}$ G \cite{Ferrer:2010wz}. In previous works, the thermodynamical properties of the SQM in the presence of strong magnetic fields and at zero temperature, within the MIT Bag model formalism, have been studied \cite{Chakrabarty:1996te, Chaichian:1999gd, Martinez:2003dz, PerezMartinez:2007zz, Felipe:2008cm}. In these references, the anisotropy produced by strong magnetic fields on the pressure components of the stress-energy tensor have been considered, due to the lost of the rotational symmetry introduced by the field.

The scope of this paper is to study the SQM in $\beta-$equilbrium, charge neutrality and constant baryon density, under the presence of a strong magnetic field and warm temperatures within the MIT Bag model formalism. The effect of the temperature and magnetic field on the particle fractions, magnetic properties, pressure anisotropy, equation of state (hereafter EoS) and stability are carefully studied numerically and correspondingly dicussed. A recent work \cite{Menezes:2011zz} has studied the magnetized SQM (hereafter MSQM) at finite temperature in the framework of the NJL model.

The paper is organized in III sections. Section I is devoted to discuss the energy spectrum per particle and the thermodynamical bulk properties of the MSQM $T\neq 0$. The Section II is dedicated to present the results of the numerical computations of the particle fractions, magnetization and spin polarization, the energy per baryon and stability conditions, the pressure anisotropy and EoS, and finally, the Mass-Radius (MR) relation for stars made of MSQM. In section III the conclusions will be presented.


\section{Thermodynamical properties of warm magnetized strange matter}

The MIT Bag model has been widely used to describe phenomenologically SQM at relatively low baryon number as in the case of strangelets \cite{Farhi:1984qu, Berger:1986ps, Gilson:1993zs, Heiselberg:1993dc, Felipe:2012}. It has been shown that surface and curvature phenomena are indeed important and should be considered, but for  large baryon densities, these corrections play almost no role when compared with the bulk properties, which has been widely studied in order to get a more comprehensive understanding of the interior of QSs. On the other hand, to consider the effects of the magnetic field on the thermodynamical properties of the MSQM, the energy spectrum per particle has to be provided. By assuming a local volume $V$ (large enough to apply the thermodynamical limit) enclosing the SQM and under the action of a constant and homogeneous magnetic field $\mathcal{B}$, pointing in the $z$-direction, the energy levels are quantized in Landau levels in the plane perpendicular to $\mathcal{B}$, so that the energy spectrum is \cite{Canuto:1969cs, Felipe:2007vb} (and references therein):
\begin{equation}\label{espectroquarks}
E_{p,f}^{\nu,\eta}=\sqrt{p_{z}^{2}+p_{\perp f}^{\;2}+m_{f}^{2}}\,,
\end{equation}
\noindent being
\begin{equation}\label{pperp}
p_{\perp f}=m_{f}\sqrt{\frac{\mathcal{B}}{\mathcal{B}_{f}^{c}}\bigl(2n-\eta+1\bigr)}\,,\quad
\mathcal{B}_{f}^{c}=\frac{m_{f}^{2}}{q_{f}},
\end{equation}
\noindent and $m_{f}$ are the electron and quark masses, $f=(e,u,d,s)$. The integer quantity $n$ indexes the Landau level, $\eta=\pm 1$ are the fermion spin projections onto the magnetic field's direction, $\mathcal{B}_{f}^{c}$ are the critical magnetic fields and $q_{f}$ denotes the corresponding electric charges. In this paper the particle masses are taken as $m_{e}=0.5$ MeV, $m_{u}=m_{d}=5$ MeV and $m_{s}=150$ MeV, hence, the critical magnetic fields for each particle are: $\mathcal{B}_{e}^{c}=4.23\times 10^{13}$ G, $\mathcal{B}_{u}^{c}=6.34\times 10^{15},$ $\mathcal{B}_{d}^{c}=1.27\times 10^{16},$ and $\mathcal{B}_{s}^{c}=1.14\times 10^{19}$ respectively.

The general expression for the bulk thermodynamic potential \cite{Chakrabarty:1996te} of a gas of electrons, quarks and gluons at finite temperature, and in the presence of an homogeneous and uniform magnetic field directed along the $z-$axis, consists of the five terms written below:
\begin{align}
\Omega=&\sum_{f=e,u,d,s}\Omega_{f\overline{f}}+\Omega_{g}+\Omega_{\gamma}+B_{\text{bag}},\label{TotalOmega}
\end{align}
\noindent where at the tree level,
\begin{align}
&\Omega_{f\overline{f}}=-\dfrac{d_{f}q_{f}\mathcal{B}}{2\pi^{2}}\sum_{\nu=0}^{\nu^{f}_{\text{max}}}[\omega_{f,\nu}^{+}(f\overline{f})+\omega_{f,\nu}^{-}(f\overline{f})],\\ \label{BOmega} 
&\omega_{f,\nu}^{\pm}(f\overline{f})=\dfrac{1}{\beta}\int_{0}^{+\infty}\ln\left[F^{+} F^{-}\right]dp_{z},\\ \label{qqeestatcontrib}
&F^{\pm}=\left[1+e^{-\beta(E_{p,f}^{\nu,\pm}\;\mp\;\mu_{f})}\right],
\end{align}
\noindent denote the statistical contributions of each particle (and their respective anti-particles) labeled by the subscript $f\;(\overline{f}).$ The degeneration factors $d_{f}$ correspond the color degrees of freedom of electrons $d_{e}=1$ and $d_{q}=3$ for quarks, and $\beta=T^{-1}$ the inverse temperature. The sum on the right hand side of \eqref{qqeestatcontrib} corresponds to the Landau levels quantization in the $p_{x},p_{y}$ plane, meanwhile the integration in the $p_{z}$ component is performed in $\omega_{f,\nu}^{\pm}(f\overline{f}).$ Finally, the functions $F^{\pm}$ are related to the Fermi-Dirac distribution with $E_{p,f}^{\nu,\pm}$ the energy given by the spectrum per particle and $\mu_{f}$ the corresponding chemical potentials. The two terms following in \eqref{TotalOmega} correspond to the statistical thermal contribution of gluons and photons, and they are given by the usual massless boson expressions
\begin{align}
\Omega_{g}=-\dfrac{d_{g}\pi^{2}}{90}\,\beta^{-4},\;\;\Omega_{\gamma}=-\dfrac{d_{\gamma}\pi^{2}}{90}\,\beta^{-4},
\end{align}
\noindent where in this case, the degeneration factors are $d_{g}=16$ (8 gluons$\times$ 2 spin projections) and $d_{\gamma}=2$ (2 spin projections). The last term in \eqref{TotalOmega} corresponds to the vacuum contributions, represented by a constant vacuum energy density $B_{\text{bag}},$ hereafter called the Bag parameter, which contains the $(\mathcal{E}_{g}^{2}+\mathcal{B}_{g}^{2}+\mathcal{B}^{2})/8\pi$ QCD and QED temperature-independent vacuum energies. Notice that since the statistical pressure of the gas satifies $P/V=-\Omega,$ then $B_{\text{bag}}$ admits the intepretation of a negative vacuum pressure that mimics the color confinement of quarks  and the Cassimir pressure inside the local volume. A more realistic description would consider $B_{\text{bag}}$ depending on the magnetic field, temperature and density, via renormalization when including higher loop corrections to $\Omega,$ which is out of the scope of the present work, so $B_{\text{bag}}$ is considered as a fixed constant.

Another feature of fermion gases are the Fermi momenta, which in the relativistic limit is given by 
\begin{align}
K_{f}=\sqrt{\mu_{f}^{2}-m_{f}^{2}},
\end{align}
\noindent where $m_{f}$ is the fermion mass and $\mu_{f}$ the gas chemical potential. The Fermi momentum has to be a real-valued quantity, thus one have to set $\mu\geq m$ for the chemical potential solutions. In the case the fermion gas is under the action of a magnetic field as described at the beginning of the section, the Fermi momenta is given instead by:
\begin{align}
K_{f}=\sqrt{\mu_{f}^{2}-p_{\perp f}^{\;2}-m_{f}^{2}},
\end{align}
\noindent and has to be a real-valued quantity. But given the chemical potential $\mu_{f},$ this condition bounds up the Landau levels accessible to each particle
present in the gas. The maximum level of Landau that each particle can be excited to is:
\begin{equation}\label{landaulevels}
\nu_{\mbox{\tiny{max}}}^{f}=I\left[\dfrac{x_{f}^{2}-1}{2\mathcal{B}/\mathcal{B}_{f}^{c}}\right],
\end{equation}
\noindent where $x_{f}=\mu_{f}/m_{f}$ is the dimensionless chemical potential per flavor and $I[x]$ the integer part of $x$.

Once known the thermodynamical potential $\Omega,$ one is able to compute the particle densities $N_{f}$, the magnetization, entropy and total energy density using:
\begin{align}
&N_{f}=-\partial\Omega/\partial\mu_{f},\\
&M\;=-\partial\Omega/\partial\mathcal{B},\\
&S\,\;\;=-\partial\Omega/\partial T,\\
&E\;\;=\Omega+TS+\sum_{f=e,u,d,s}\mu_{f}N_{f}.
\end{align}
\noindent The expressions of $N_{f},\,M,\,S,\,E$ and $\Omega$ at $T=0$ can be found in ~\cite{Felipe:2007vb}, however, in this article the full expresions at $T\neq0$ are considered. Last but not least, recall that in the presence of a strong magnetic field in the direction of the $z-$axis, the stress-energy-tensor contains anisotropic pressure components parallel $P_{||}$ and transversal $P_{\perp}$ to the field, both related by: 
\begin{align}
&P_{||}=-\Omega,\\
&P_{\perp}=P_{||}-M\mathcal{B}.
\end{align}  
In the following section, all these magnitudes will be evaluated for the MSQM in $\beta-$equilibrium, electric charge neutrality and baryonic number conservation, at finite temperatures.

\section{Magnetized strange quark matter at $T\neq0$}

This section is devoted to study, via numerical calculations, all the relevant thermodynamical properties of the MSQM at warm temperatures, i.e. the chemical potentials and particle fractions, the magnetization, the energy density and the energy per baryon, the EoS and the pressure anisotropy, and finally the mass-radius relation for magnetized QSs. To achieve this, we need four constrain equations on the quark and electron gasses, in order obtain a self-consistent solution for the intensive variables of the thermodynamical system, which in this article, are taken the chemical potentials $\mu_e,\,\mu_u,\,\mu_d$ and $\mu_s$. At the same time, these constrain equations must match astrophysical observables in order to give a physical explanation to the obtained solutions and derived quantities. There is a vast literature where one can find the most common constrain equations \cite{Chakrabarty:1996te, Farhi:1984qu, Felipe:2007vb, Felipe:2008cm, Felipe:2010vr, Felipe:2012}, which are: $\beta-$equilibrium conditions, electric charge neutrality and baryonic number conservation. Mathematically they can be expressed by:

\begin{align}
&\mu_{u}+\mu_{e}=\mu_{d},\;\;\mu_{d}=\mu_{s},\label{betaequil}\\
&2N_{u}-N_{d}-N_{s}=3N_{e},\label{electriccharge}\\
&N_{u}+N_{d}+N_{s}=3N_{B}.\label{baryondensityconservation}
\end{align}

The first condition, given by Eq.\eqref{betaequil}, means that if SQM really exists in the core of NSs or in SQSs, then the appearance of $s$ quarks will be through weak interaction processes, where it is assumed that the relaxation time is long enough to safely neglect the neutrino-antineutrino contributions \cite{Hatsuda:1987ck, Sato:1987rd}; once this happens, the equilibrium among quarks-antiquarks and electrons-positrons will be dynamically established. The second condition, ensured by Eq.\eqref{electriccharge}, is the local electric charge neutrality, that enhances the unobservable charged astrophysical objects. The last condition is the local baryon density conservation, given by Eq.\eqref{baryondensityconservation}, that enhances the color confinement of quarks within the MIT Bag Model, where $N_{B}$ is the baryon density. Once solved this four equations for four the chemical potentials, one is able to evaluate the above mentioned thermodynamical quantities and study their behavior as a function of any of the three parameters $(\mathcal{B}, N_{B}, T)$ leaving constant the other two. Hereafter this will be the approach in this paper. Last but not least, and as already mentioned in the previous section, in Ref.\cite{Felipe:2007vb} were already studied the thermodynamical properties of the MSQM at $T=0$ in the MIT Bag Model framework. In this section I compare those results with the warm temperature cases.

\bigskip

\subsection{Chemical potentials and particle fractions}

\begin{figure}[h!t]
\centering
\includegraphics[width=0.45\textwidth]{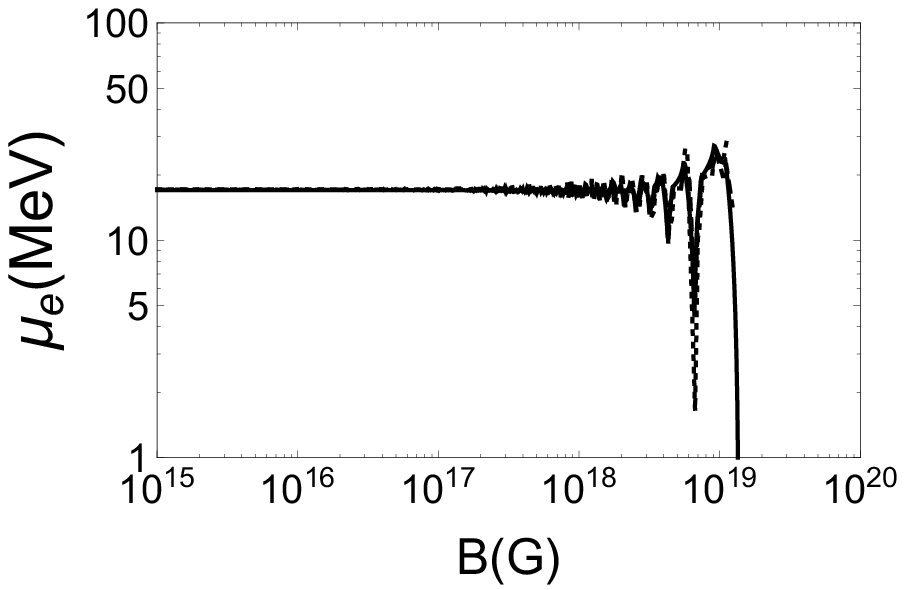}\\
\includegraphics[width=0.45\textwidth]{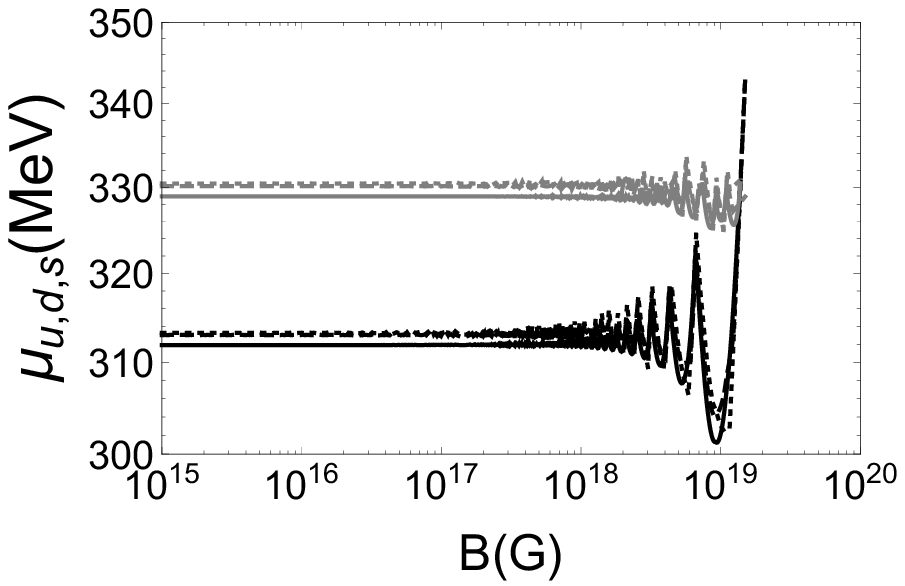}\\
\caption{Magnetic field dependence of the chemical potentials of electrons (upper panel) at a fixed baryon density $N_{B}=2.5\times N_{0}$ and temperatures $T=0,\, 15$ and $30$ MeV (solid, dashed and dotted lines). In the lower panel are shown the chemical potentials of u-quarks (black lines) and d, s-quarks (gray lines) for the same baryon density and temperatures.}\label{Fig0012}
\end{figure}

As already discussed, with the aid of Eq.\eqref{betaequil}, Eq.\eqref{electriccharge} and Eq.\eqref{baryondensityconservation}, one can obtain the chemical potentials of all the particle flavours involved in the SQM, which are shown in Fig.\eqref{Fig0012} as a function of the external magnetic field $\mathcal{B}$, for fixed baryon density $N_{B}=2.5\times N_{0}$ and temperatures $T=0, 15, 30$ MeV respectively.

In all cases, including warm temperatures, the chemical potentials behave almost constant for $\mathcal{B}\leq10^{17}$ G. As $\mathcal{B}$ continues increasing from this value on, the chemical potentials begin to show oscillations, strengthening in amplitude for higher values of the external field, until reaching a saturation value of about $\mathcal{B}_{\text{s}}\sim1.37\times10^{19}$ G, where the electrons chemical potential drops sharply to zero and becomes negative. Positive solutions for the chemical potentials are found for higher fields than $\mathcal{B}_{\text{crit}}$, as shown in Ref.\cite{Felipe:2007vb}. In this article, no ultra-strong magnetic fields are considered since one needs to include higher quantum corrections, plus these fields are unlikely to appear in COs or in particle accelerators. Moreover, for warm temperatures and $\mathcal{B}\leq10^{17}$ G, it is observed a slightly increase in the chemical potentials, which is more noticeable for the $u$ and $d, s$ quarks than for electrons as also shown in Fig.\eqref{Fig0012}, and in the strong field regime one can observe the same oscillations and the same behavior of electrons chemical potentials as described above.

\begin{figure}[h!t]
\centering
\includegraphics[width=0.225\textwidth]{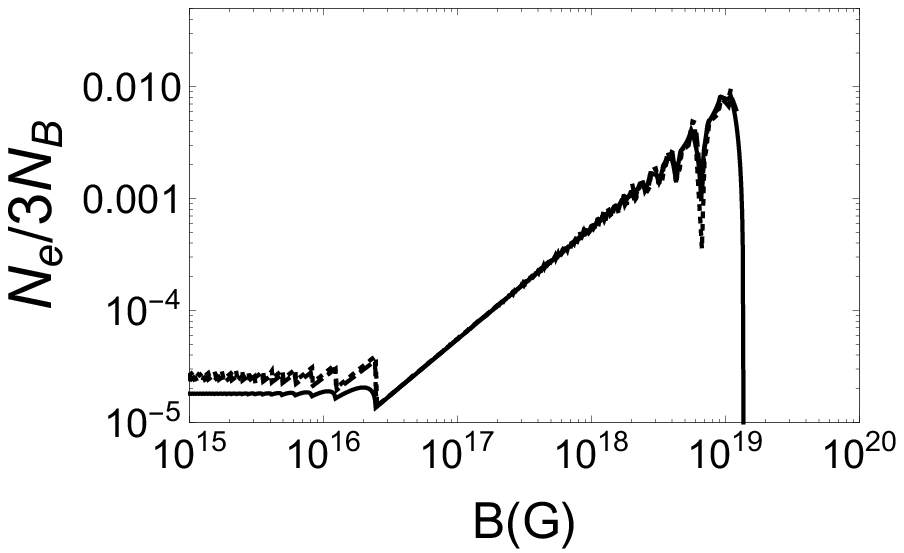} \includegraphics[width=0.225\textwidth]{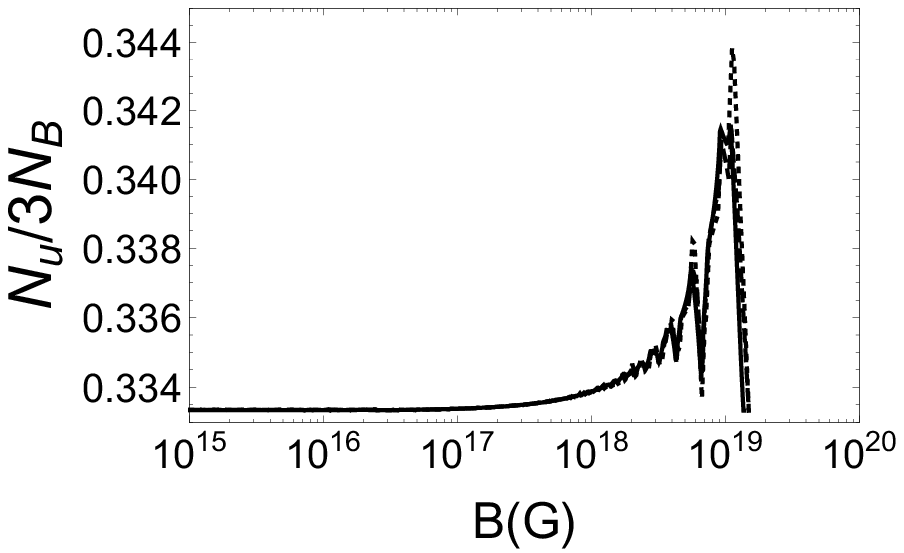}\\
\includegraphics[width=0.225\textwidth]{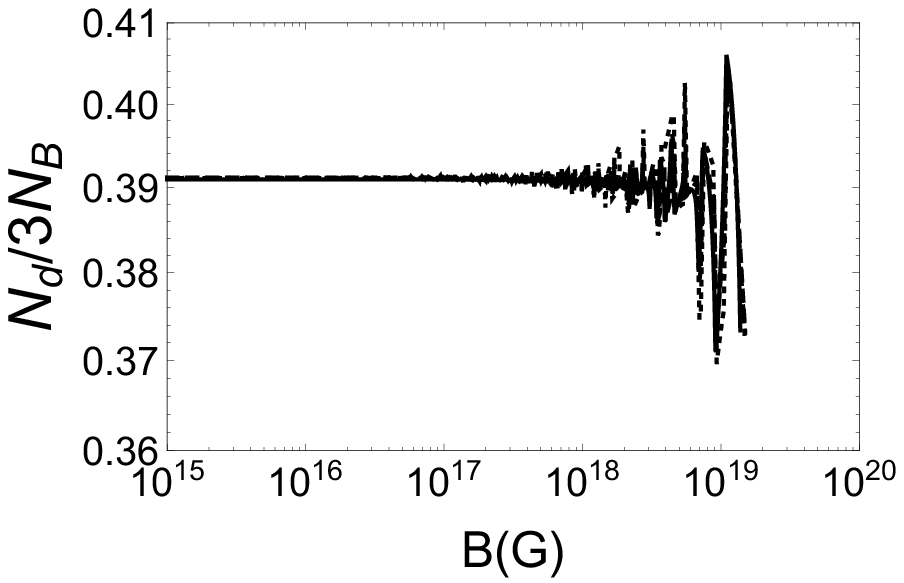} \includegraphics[width=0.225\textwidth]{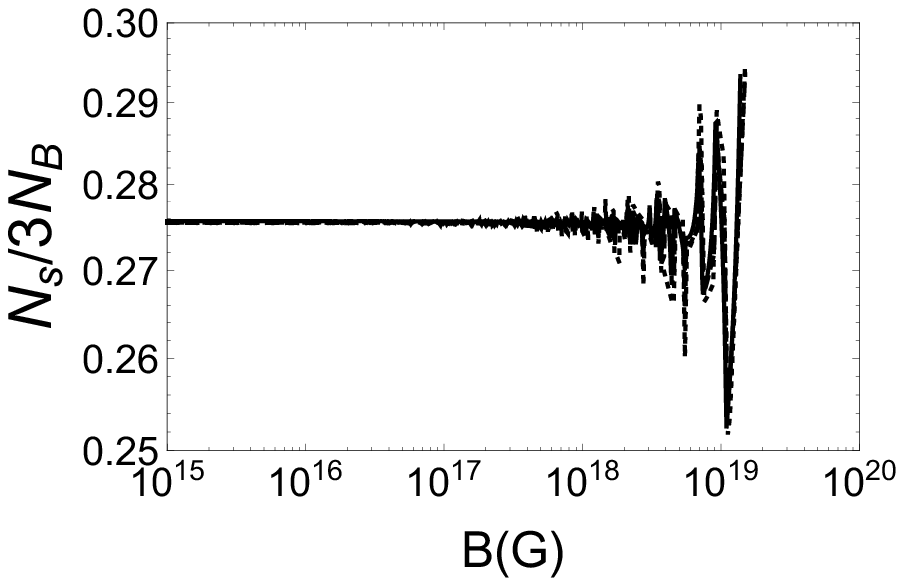}\\
\caption{Magnetic field dependence of the particle fractions $N_{i}/3N_{B}$ for electrons (upper left panel), u-quarks (upper right panel), d-quarks (lower left panel) and s-quarks (lower right panel), at a fixed baryon density $N_{B}=2.5\times N_{0}$ and temperatures $T=0,\, 15$ and $30$ MeV (solid, dashed and dotted lines).}\label{Fig003456}
\end{figure}

In Fig.\eqref{Fig003456} are shown the particle fractions $N_{i}/3N_{B}$ as a function of $\mathcal{B}$ for a fixed baryon density $N_{B}=2.5\times N_{0}$ and temperatures $T=0, 15, 30$ MeV respectively. At $T=0$, electrons are distributed filling a Fermi sea in momentum space up to a maximum Landau level $\nu_{\mbox{\tiny{max}}}^{f}$, which depends proportionally to $\mu_{e}^{2}$ and inversely proportional to $\mathcal{B}$ as shown in Eq.\eqref{landaulevels}. By increasing $\mathcal{B}$ beyond $\sim10^{19}$ G, electrons are forced to populate the ground level since they have the lowest critical field $\mathcal{B}_{e}^{c}$, and due to the Pauli exclusion principle, the amount of these will be decimated causing an abrupt decrease in the electron fraction as shown in the upper left panel of Fig.\eqref{Fig003456}. In parallel, the u-quarks fraction decreases as well to compensate the decresing of the negative charge. A vanishing electron chemical potential leads to the equality of the chemical potentials of quarks $\mu_{u}=\mu_{d}=\mu_{s}$ from Eq.(\ref{betaequil}), and therefore the local electric charge neutrality condition is not longer ensured and no more solutions to the equations are found. On the other hand, the fraction of $d$ quarks decreases due to the large value of $\mathcal{B}/\mathcal{B}_{d}^{c}$ compared to $\mathcal{B}/\mathcal{B}_{s}^{c}$ and the Pauli exclusion principle, thus the fraction of $s$ quarks will increase, in accordance with the fixed total baryonic density Eq.(\ref{baryondensityconservation}). Since the maximum Landau level of each gas flavour does not depend on the temperature, the same reasoning applies at warm temperatures.

For $\mathcal{B}\leq10^{17}$ G the particle fractions are almost constant, however, as the magnetic field starts to increase, the electron and u-quark fractions increase as well due to the local conservation of the electric charge, while the fractions of $d$ and $s$ quarks will balance in order to keep ensuring the neutrality. The temperature corrections don't make important contributions except to electrons at low values of the magnetic field.

Last but not least, as $N_{e}/3N_{B}$ has a discontinuity with $\mathcal{B}$ at around $\sim1.37\times10^{19}$ G, the total density $N_{MSQM}=N_{e}+3N_{B}$ suffers from the same effect, indicating a possible first order phase transition from the MSQM to a nucleated phase like strangelets, where the surface effects play a weighting role, as it was studied in \cite{Felipe:2012} and the references therein.

\bigskip

\begin{figure}[h!t]
\centering
\includegraphics[width=0.45\textwidth]{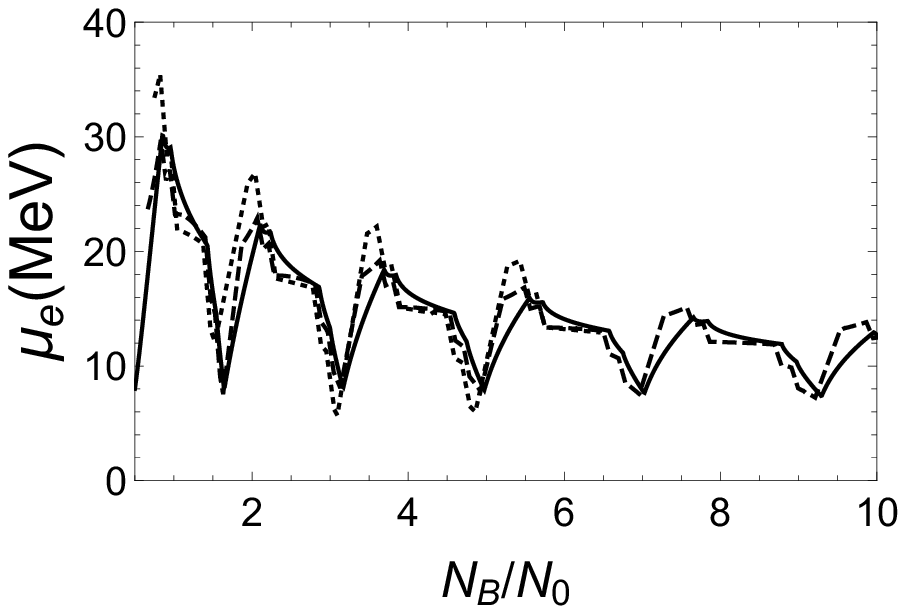}\\
\includegraphics[width=0.45\textwidth]{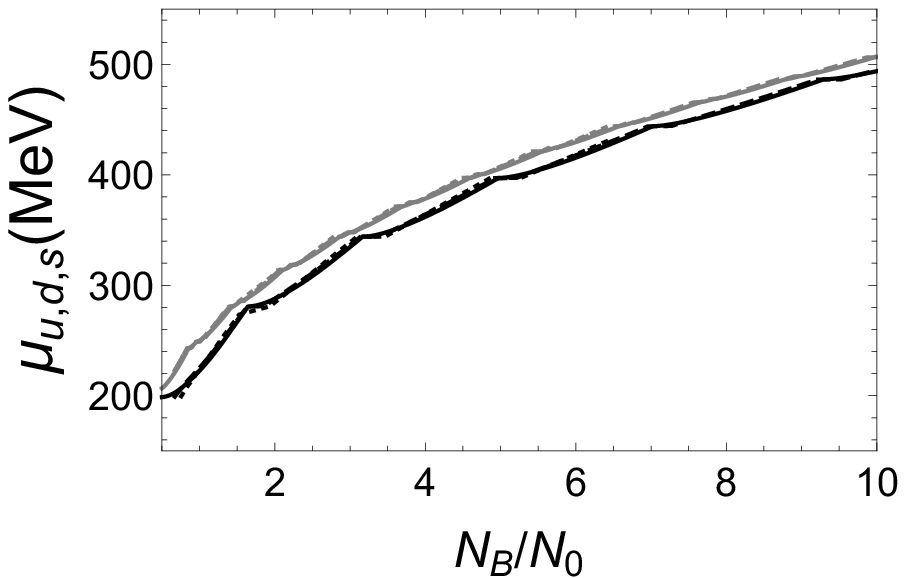}\\
\caption{Density dependence of the chemical potentials of electrons (upper panel) at a fixed external magnetic field $\mathcal{B}=5\times10^{18}$ G and temperatures $T=0,\, 15$ and $30$ MeV (solid, dashed and dotted lines). In the lower panel are shown the chemical potentials of $u-$quarks (black lines) and $d$, $s-$quarks (gray lines) for the same baryon density and temperatures.}\label{Fig0078}
\end{figure}

For a fixed magnetic field $\mathcal{B}=5\times10^{18}$ G and temperatures $T=0,\,15,\,30$ MeV, Eq.\eqref{betaequil}, Eq.\eqref{electriccharge} and Eq.\eqref{baryondensityconservation} again will be solved as a function of the baryon density $N_{B}.$ The corresponding chemical potential solutions for electrons (upper panel) and quarks (lower pannel) are shown in Fig.\eqref{Fig0078}. The electron chemical potential shows a sign of damped oscillation with the density, while the quark chemical potentials always inrease showing some periodic bumps, more preceivables for u-quarks, at the same densities as the electrons.

\begin{figure}[h!t]
\centering
\includegraphics[width=0.225\textwidth]{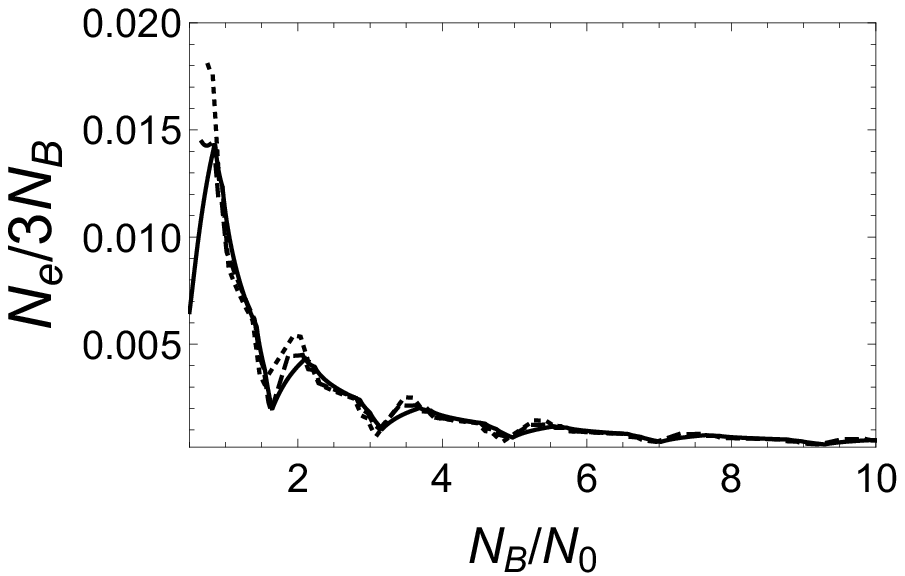} \includegraphics[width=0.225\textwidth]{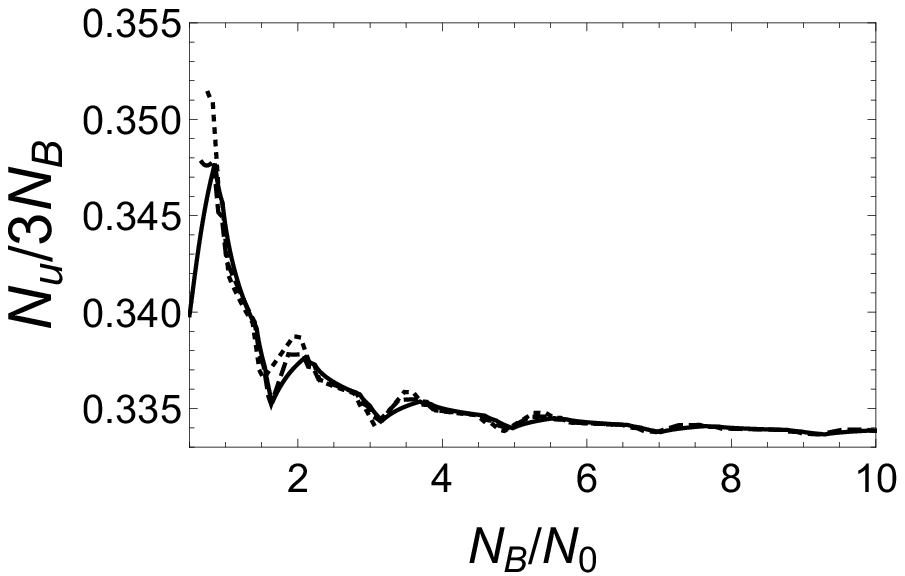}\\
\includegraphics[width=0.225\textwidth]{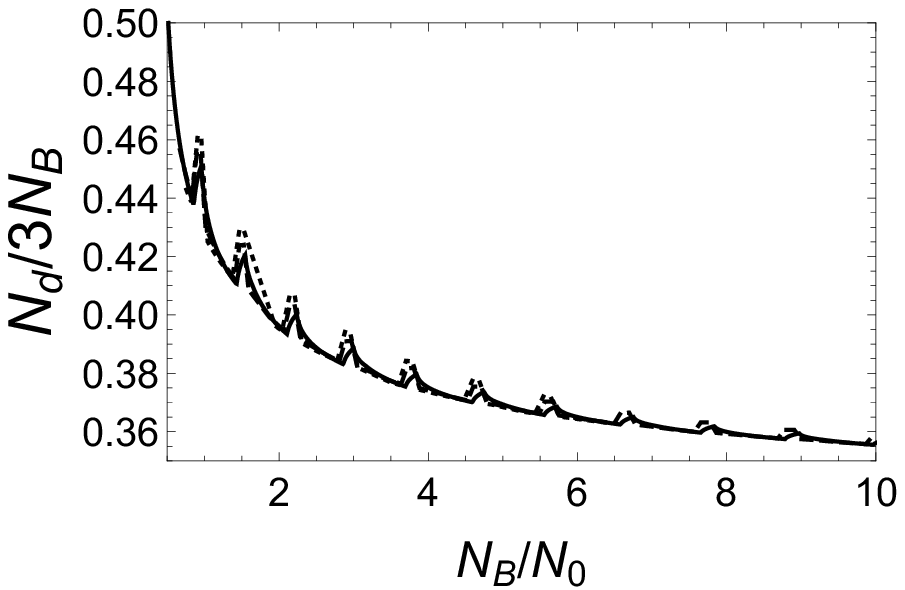} \includegraphics[width=0.225\textwidth]{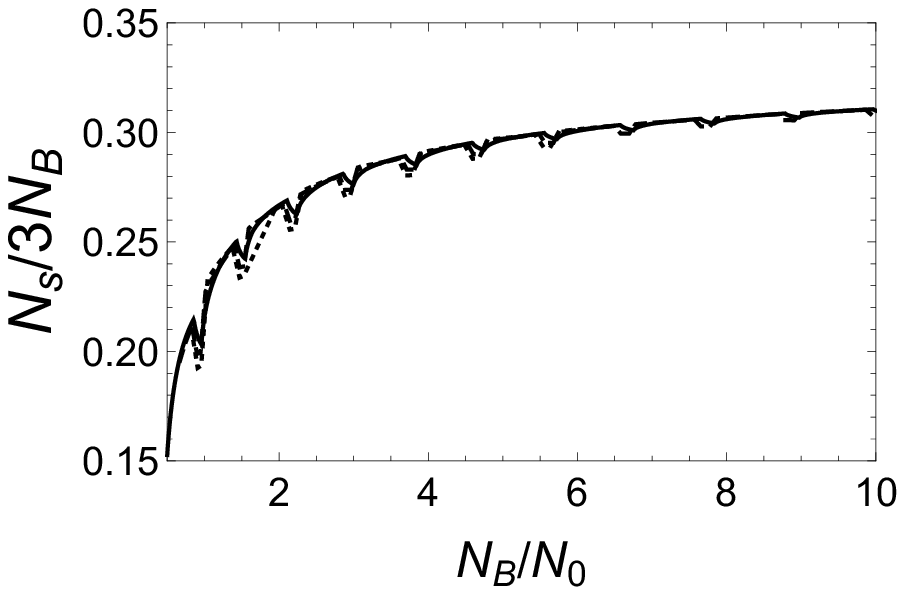}\\
\caption{Density dependence of the particle fractions $N_{i}/3N_{B}$ of electrons (upper left panel), $u-$quarks (upper right panel), $d-$quarks (lower left panel) and $s-$quarks (lower right panel), at a fixed external magnetic field $\mathcal{B}=5\times10^{18}$ G and temperatures $T=0,\, 15$ and $30$ MeV (solid, dashed and dotted lines).}\label{Fig009101112}
\end{figure}

At very low baryon densities, as shown in Fig.\eqref{Fig009101112} (lower right panel), the $s-$quarks fraction is negligible compared to the fractions of electrons, $u$ and $d-$quarks, showing that at very low densities, the magnetized nuclear matter is the state predominant in this environment. As a slow compression process starts to happen at fixed $\mathcal{B}$ and $T$, slow enough such that neutrinos have already left the local volume, the MSQM starts to appear by converting $d$ and $u-$quarks into $s$ and $u-$quarks by inverse $\beta-$processes. The electric charge neutrality ensures that for each $u-$quark there are at least two $d-$quarks, therefore, one of these two $d-$quarks is used to produce a $u-$quarks and an electron via the reaction $d\rightarrow u+e^{-}+\overline{\nu_{e}}$, thus increasing the $u-$quark and electron fractions. The other available $d-$quark scatters with the newly formed $u-$quark to produce a $s-$quark and a $u-$quark via $d+u \leftrightarrow u+s.$ As one can see from Fig.\eqref{Fig009101112}, once arrived to a baryon density of around $0.86N_{0}$, the electrons, $u$ and $s-$quarks production reaches a local maximum while the $d-$quark fraction reaches a local minimum at $0.96N_{0}$. If one keeps compressing beyond $0.86N_{0},$ then the process reverses, decreasing the electron, $u$ and $s-$quark fractions, increasing the $d-$quarks fraction via the reactions $u+e^{-}\leftrightarrow d+\nu_{e}$ and $s + u \leftrightarrow u + d$, however, the amount of $d-$quarks is not enough to account for the $s-$quarks production, therefore the over-production of electrons and $u-$quarks in the first part of the cycle have to account for more $s-$quarks via $u+e^{-} \leftrightarrow s+\nu_{e},$ which is the reason why the electron and $u-$quark fractions keep decreasing. The cycle repeats periodically with the over-production of $s-$quarks and consuming almost all the electrons availables during the process, such that if one keeps compressing, reaching densities beyond $10\times N_{0}$, every quark fraction $N_{u}/3N_{B},\,N_{d}/3N_{B}$ and $\,N_{s}/3N_{B}$ approaches to $1/3$, arriving to a high density phase with no electrons and $N_{u}=N_{d}=N_{s}$ like in the Magnetic Color-Flavour-Locked (MCFL) superconducting phase \cite{Alford:1999pb, Gorbar:2000ms, Ferrer:2006vw, Ferrer:2006ie, Ferrer:2007iw, Fukushima:2007fc, Alford:2010qf}, conjectured by the QCD phase diagram \cite{Schafer:1999fe, Alford:1998mk, Shovkovy:1999mr, Alford:2001zr, Alford:2007xm, Alford:2004pf}. Warm temperatures don't affect much the particle fractions.

\subsection{Energy density and MSQM stability}

\begin{figure}[h!t]
\centering
\includegraphics[width=0.45\textwidth]{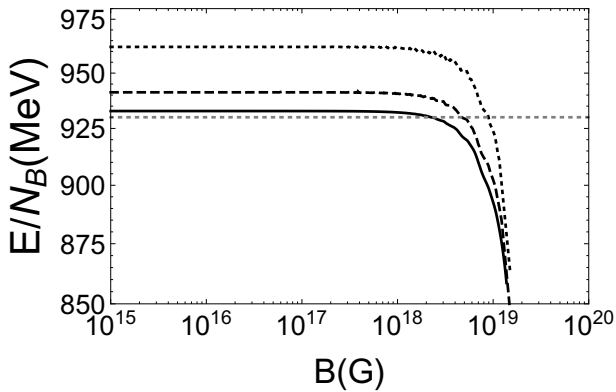}\\
\caption{Magnetic field dependence of the energy per baryon at a fixed baryon density $N_{B}=2.5\times N_{0}$, temperatures $T=0,\, 15$ and $30$ MeV (solid, dashed and dotted lines), and $B_{\rm bag}=75$~MeV~fm$^{-3}$.}\label{Fig0020}
\end{figure}

In the previous subsection were studied the solutions of Eq.\eqref{betaequil}, Eq.\eqref{electriccharge} and Eq.\eqref{baryondensityconservation} and the corresponding particle fractions in terms of the external magnetic field and the baryon densities. In this subsection, the stability condition of the MSQM, measured by the energy per baryon $E/N_{B}$ will be studied. In Fig.\eqref{Fig0020} is shown the $E/N_{B}$ as a function of the external magnetic field, for a fixed baryon density $N_{B}=2.5\times N_{0},$ temperatures $T=0,\,15,\,30$ MeV (continuous line, dashed line, dotted line) and a Bag constant of $B_{\rm bag}=75$~MeV~fm$^{-3}$ respectively. 

\noindent The gray dotted line corresponds to the $E/N_{B}=930$ MeV of the iron isotope $^{56}$Fe which is the most stable isotope in Nature. The magnetic field enhances the stability for the choosen Bag constant, baryon density and warm temperatures till the saturation field $\mathcal{B}_{\text{s}}\sim1.37\times10^{19}$ above which there are no longer solutions, as discussed in the previous subsection. At $T=0,\, 15,\, 30$ MeV, the stability fields are $\mathcal{B}_{0}=2.25\times10^{18}$ G, $4.70\times10^{18}$ G and $9.30\times10^{18}$ G respectively, which means that for fields below these stability fields, MSQM could be in a metastable phase, and beyond, the magnetic field stabilizes the MSQM against any decay mechanism for the given densities, temperatures and Bag constant. This results however, depend strongly on the Bag constant value. On the other hand, for $\mathcal{B}\leq1.\times10^{18}$ G, the energy per baryon remains almost constant and the temperature tends to increase the $E/N_{B}$ as expected from the new thermal degrees of freedom.

\begin{figure}[h!t]
\centering
\includegraphics[width=0.45\textwidth]{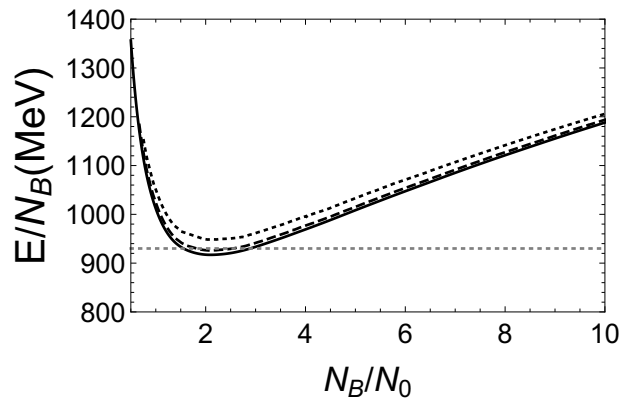}\\
\caption{Density dependence of the energy per baryon at a fixed value of the magnetic field $\mathcal{B}=5\times10^{18}$ G, temperatures $T=0,\, 15$ and $30$ MeV (solid, dashed and dotted lines) and $B_{\rm bag}=75$~MeV~fm$^{-3}$.}\label{Fig0021}
\end{figure}

In Fig.\eqref{Fig0021} it is shown the dependence of the energy per baryon with the baryon density for $\mathcal{B}=5\times 10^{18}$~G, a Bag constant of $B_{\rm bag}=75$~MeV~fm$^{-3}$ and temperatures of $T=0, 15, 30$~MeV respectively. Following the analysis of the particle fractions in the preceeding subsection, the MNM gains stability by producing a new heavy flavour and thus converting to MSQM, which if it keeps compressing, the energy per baryon goes below $930$ MeV entering in the stability zones $1.56\times N_{0}\leq N_{B}\leq 2.89\times N_{0}$ for $T=0$ MeV and $1.70\times N_{0}\leq N_{B}\leq 2.54\times N_{0}$ for $T=15$ MeV, reaching both their global minima at $N_{B}=2.1\times N_{0}$. This means that for the settled environment, the MSQM could be absolutely stable and long lasting in Nature in the correspondent range of baryonic densities. 

\begin{figure}[h!t]
\centering
\includegraphics[width=0.45\textwidth]{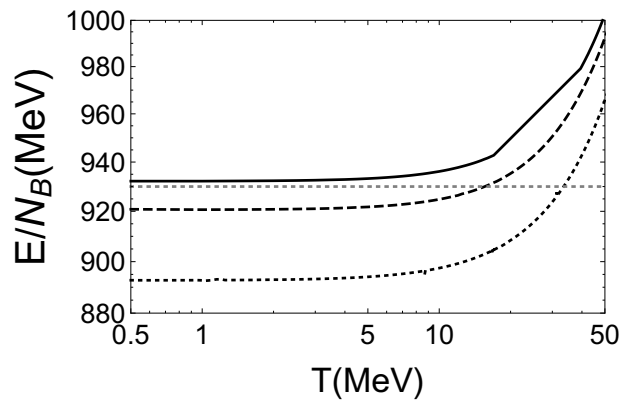}\\
\caption{Temperature dependence of the energy per baryon at fixed values of the magnetic field $\mathcal{B}=1\times10^{18}$ G (continuous line), $\mathcal{B}=5\times10^{18}$ G (dashed line) and $\mathcal{B}=1\times10^{19}$ G (dotted line), fixed baryonic density $N_{B}=2.5\times N_{0}$, and $B_{\rm bag}=75$~MeV~fm$^{-3}$.}\label{Fig0022}
\end{figure}

In Fig.\eqref{Fig0022} it is shown the expected increasing behavior of the energy per baryon with the temperature due to the thermal motion of quarks, electrons and the contributions of gluons. In addition, the range of temperatures in which the MSQM is absolutely stable increases with the magnetic field.

\subsection{Magnetic field response}

\bigskip

\begin{figure}[h!t]
\includegraphics[width=0.45\textwidth]{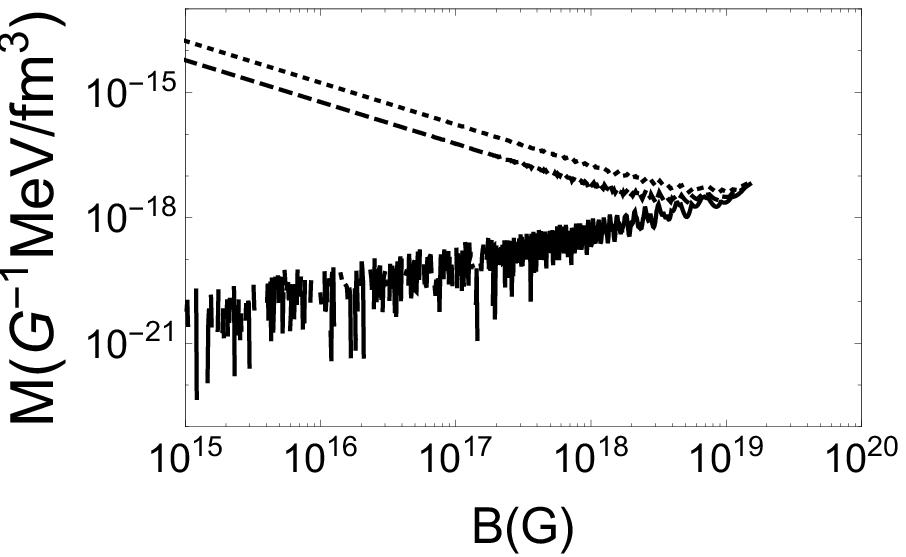}\\
\includegraphics[width=0.45\textwidth]{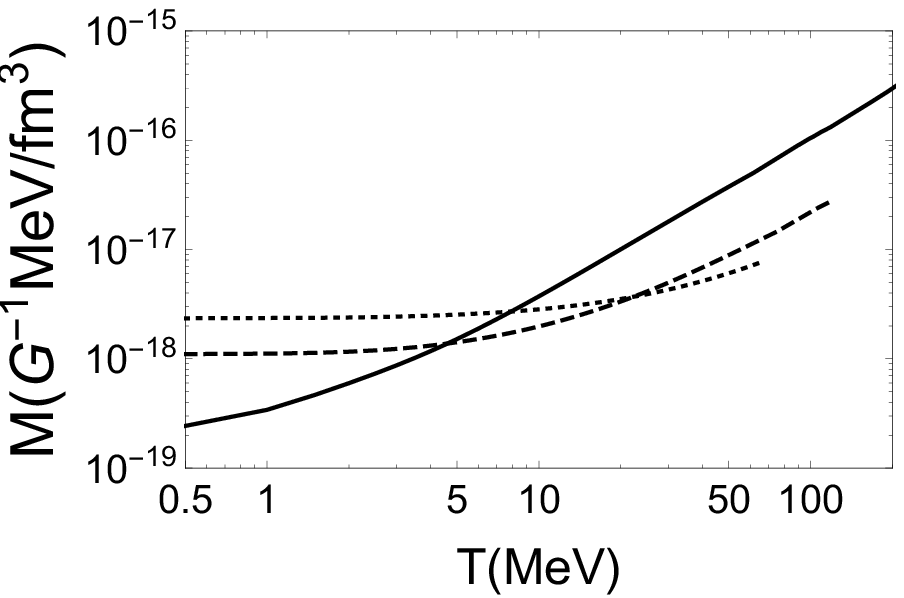}\\
\caption{Magnetic field dependence of the magnetization (upper panel) at fixed $N_{b}=2.5\times N_{0}$, and $T=0, 15, 30$~MeV (continuous, dashed and dotted lines). Temperature dependence of the magnetization (lower panel) at fixed $N_{b}=2.5\times N_{0}$, and $\mathcal{B}=1\times 10^{18},\,5\times 10^{18}$ and $1\times 10^{19}$ G (continuous, dashed and dotted lines).}\label{Fig02324}
\end{figure}

The magnetization of the MSQM is shown in the upper panel of Fig.\eqref{Fig02324} as a function of the external magnetic field at fixed baryonic density $N_{B}=2.5\times N_{0}$ and temperatures $T=0, 15, 30$ MeV respectively. At zero temperature, the MSQM shows the typical Haas-van Alphen oscillations as the response of the MSQM to the magnetic field, with the over all positive magnetization increasing with the applied field in the range of solutions of the equations Eq.\eqref{betaequil}, Eq.\eqref{electriccharge} and Eq.\eqref{baryondensityconservation} respectively. The finite temperature magnetization shows an increasing behavior with the temperature and decreasing with the magnetic field as shown as well in the lower panel of Fig.\eqref{Fig02324}. This behavior was already discussed in Ref.\cite{Skobelev:2012tc, Daicic:1994td} and references therein.

\begin{figure}[h!t]
\includegraphics[width=0.45\textwidth]{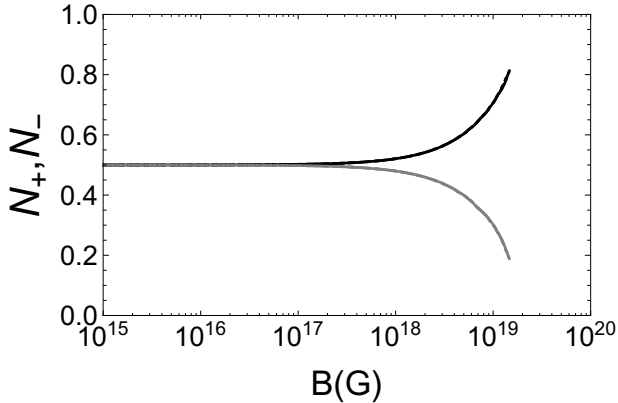}\\
\caption{Magnetic field dependence of the spin polarization (black lines $N_{+}$, gray lines $N_{-}$) at fixed $N_{b}=2.5\times N_{0}$, and $T=0, 15, 30$~MeV (continuous, dashed and dotted lines).}\label{Fig02526}
\end{figure}

The total particle spin polarization is shown in the upper panel of Fig.\eqref{Fig02526} as a function of the magnetic field. The black lines correspond to the total number of particles with spin in the direction of the magnetic field $N_{+}$ and the gray lines correspond to the ones with spin flipped against the magnetic field $N_{-}$. An increase of the magnetic field produces the increase of the polarization being more perceivable for fields above $10^{17}$ G, however, a total polarization is never achieved since beyond $\mathcal{B}_{\text{s}}\sim1.37\times10^{19}$ G there are no longer solutions for the chemical potentials. On the other hand, warm temperatures seem not to make any significant contribution to the spin polarization.

\bigskip

\subsection{Pressure anisotropy and EoS}

\begin{figure}[h!t]
\includegraphics[width=0.45\textwidth]{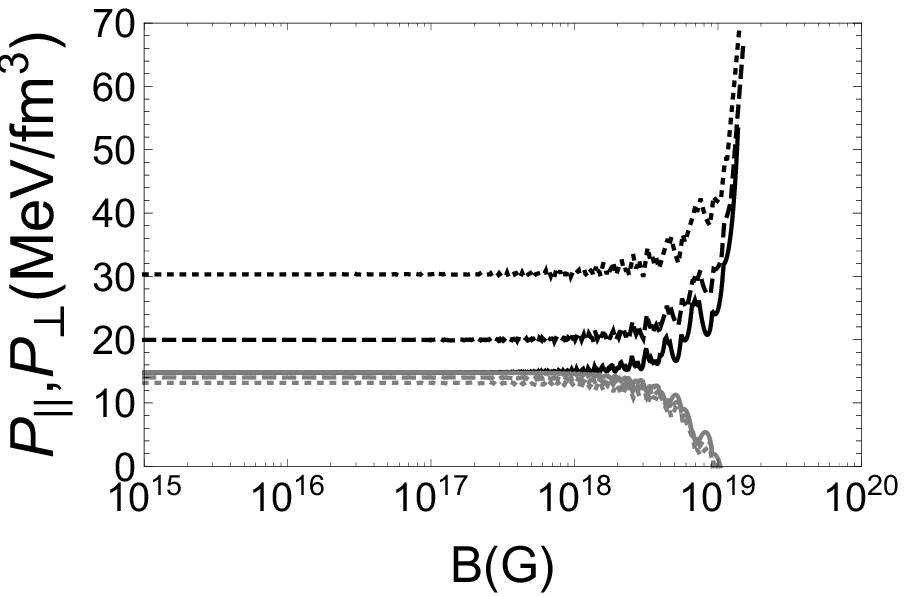}\\
\includegraphics[width=0.45\textwidth]{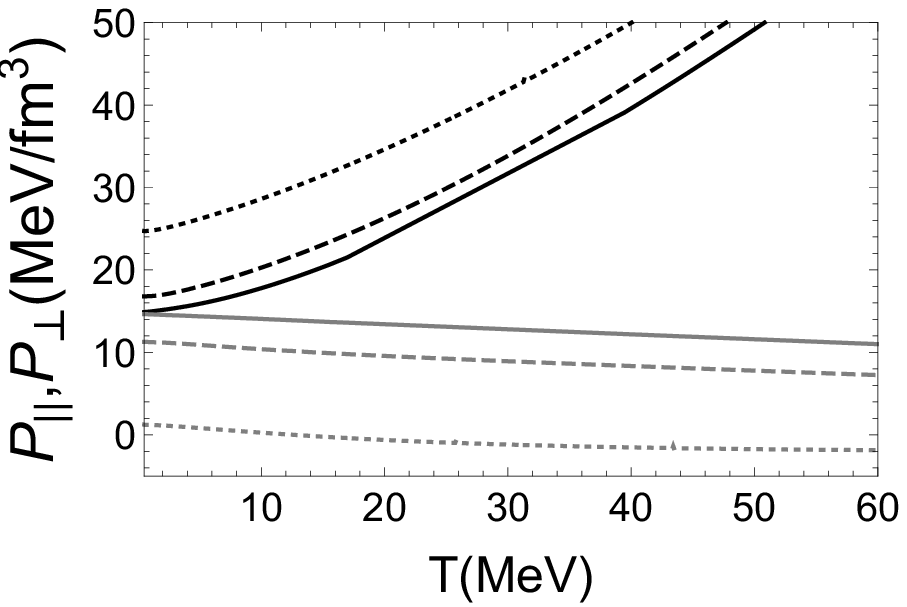}\\
\caption{Magnetic field dependence (upper panel) of the parallel (black line) and transversal pressures (gray line) at fixed $N_{B}=2.5\times N_{0}$, $B_{\rm bag}=75$~MeV~fm$^{-3}$ and $T=0, 15, 30$~MeV (continuous, dashed and dotted lines). Temperature dependence (upper panel) of the parallel (black line) and transversal (gray line) pressures at fixed $N_{B}=2.5\times N_{0}$, $B_{\rm bag}=75$~MeV~fm$^{-3}$ and $\mathcal{B}=1\times10^{18},\; 5\times10^{18}$ G and $1\times10^{19}$~G (continuous, dashed and dotted lines).}\label{Fig02728}
\end{figure}

The presence of a strong magnetic field breaks the spherical symmetry with the appearance of an anisotropy in the pressure components of the stress-energy tensor. These two pressures, i.e. parallel $P_{||}$ and transversal $P_{\perp}$ pressures are shown in the upper panel of Fig.\eqref{Fig02728} as a function of the magnetic field, at fixed density $N_{B}=2.5\times N_{0}$, $B_{\rm bag}=75$~MeV~fm$^{-3}$ and $T=0,\, 15,\, 30$ MeV (continuous, dashed and dotted lines). At $T=0$ MeV the anisotropy manifests for fields larger than $10^{17}$ G. At warm temperatures, since the magnetization of the MSQM increases with $T$ at a fixed density, the anisotropy persists even at lower values of the external field. However, and almost independently of the temperature, once reached the saturation field $\mathcal{B}_{\text{s}}\sim1.37\times10^{19}$ G, the local volume collapses equatorially since at $\mathcal{B}_{\text{s}},$ $P_{\perp}$ vanishes and becomes negative beyond $\mathcal{B}_{\text{s}}$. The parallel pressure $P_{||}$ always increases with $\mathcal{B}$ and $T.$

In the lower panel of Fig.\eqref{Fig02728} are shown as well $P_{||}$ and $P_{\perp}$, this time, as a function of the temperature at fixed density $N_{B}=2.5\times N_{0}$, $B_{\rm bag}=75$~MeV~fm$^{-3}$ and $\mathcal{B}=1\times10^{18},\, 5\times10^{18},\, 1\times10^{19}$ G (continuous, dashed and dotted lines). The temperature enhances the anisotropy and an equatorial collapse is forseen only for $\mathcal{B}=1\times10^{19}$ G and above, in the range of temperatures analyzed. 

\begin{figure}[h!t]
\includegraphics[width=0.45\textwidth]{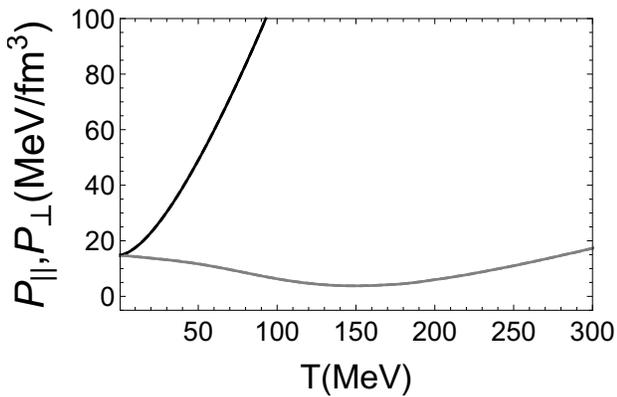}\\
\caption{Temperature dependence of the parallel (black line) and transversal (gray line) pressures at fixed $N_{B}=2.5\times N_{0}$, $B_{\rm bag}=75$~MeV~fm$^{-3}$ and $\mathcal{B}=1\times10^{16}$ G.}\label{Fig028a}
\end{figure}

\noindent Depending on the baryon density, one should expect a minimum of $P_{\perp}$ at a characteristic temperature $T_{c}$ in such a way the transversal pressure starts to increase beyond $T_{c}$; this is shown in Fig.\eqref{Fig028a} for $\mathcal{B}=1\times 10^{16}$ G. The parallel pressure $P_{||}$ always increases with $\mathcal{B}$ and $T$ as expected.

\begin{figure}[h!t]
\includegraphics[width=0.45\textwidth]{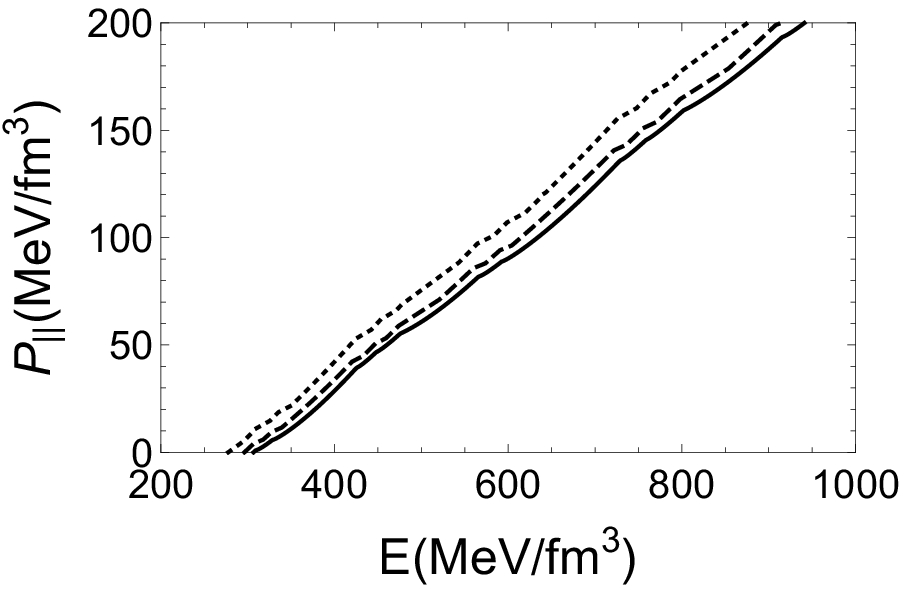}\\
\caption{EoS for the MSQM at fixed magnetic field $\mathcal{B}=5\times10^{18}$ G, $B_{\rm bag}=75$~MeV~fm$^{-3}$ and $T=0, 15, 30$~MeV (continuous, dashed and dotted lines).}\label{Fig029}
\end{figure}

The EoS for a fixed  magnetic field $\mathcal{B}=5\times10^{18}$ G, $B_{\rm bag}=75$~MeV~fm$^{-3}$ and $T=0, 15, 30$~MeV (continuous, dashed and dotted lines), are plotted in Fig.\eqref{Fig029}. The increase of the pressure with the energy density is almost linear, with the corresponding speeds of sound $0.566c,\,0.571c,$ and $0.581c$ for $T=0, 15$ and $30$ MeV respectively.

\begin{figure}[h!t]
\includegraphics[width=0.45\textwidth]{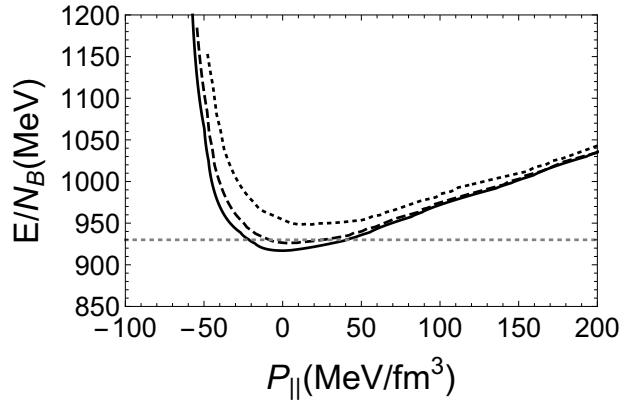}\\
\caption{Energy per baryon as a function of the parallel pressure at fixed $\mathcal{B}=5\times10^{18}$ G, $B_{\rm bag}=75$~MeV~fm$^{-3}$ and $T=0, 15, 30$~MeV (continuous, dashed and dotted lines).}\label{Fig030}
\end{figure}

The energy per baryon $E/N_{B}$ is depicted in Fig.\eqref{Fig030} once again, but this time as a function of the parallel pressure $P_{||}$, at a fixed  magnetic field $\mathcal{B}=5\times10^{18}$ G, $B_{\rm bag}=75$~MeV~fm$^{-3}$ and $T=0, 15, 30$~MeV (continuous, dashed and dotted lines). The global minimum of the $E/N_{B}$ is attained at zero parallel pressure, thus becoming once more a confirmation of the Bodmer-Witten-Terasawa conjecture about the stability of SQM in the presence of a strong magnetic field and warm temperatures.

\subsection{Mass-Radius relation for MSQM Stars}

\begin{figure}[h!t]
\includegraphics[width=0.45\textwidth]{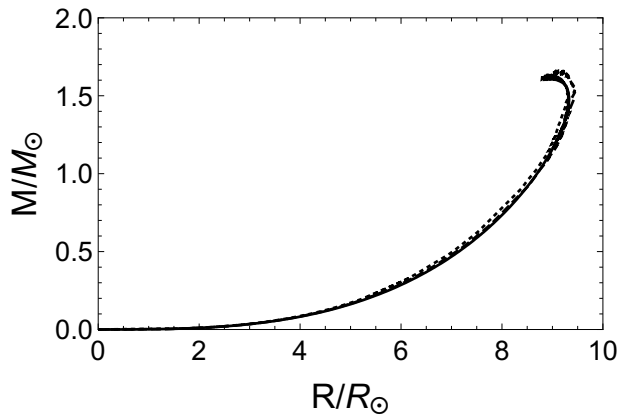}\\
\caption{Mass radius relation for magnetized strange quark stars with $\mathcal{B}=5\times 10^{18}$~G, $B_{\rm bag}=75$~MeV~fm$^{-3}$ and $T=0, 15, 30$~MeV.}\label{Fig031}
\end{figure}

For a spherical non-rotating compact object provided an EoS $F(P(r),E(r))=0$, the Toldman-Oppenheimer-Volkof equations allow the computation of the total mass $M(r)$ enclosed by a sphere of radius $r$ via:
\begin{align}
&\dfrac{1}{E}\dfrac{dM}{dr}=4\pi r^{2} G,\label{massTOV}\\
&\dfrac{1}{E}\dfrac{dP}{dr}=-\dfrac{G M}{r^{2}}\dfrac{\left(1+\dfrac{P}{E}\right)\left(1+\dfrac{4\pi r^{3}P}{M}\right)}{\left(1-\dfrac{2G M}{r}\right)}\label{pressTOV}.
\end{align}
\noindent Given the boundary conditions $M(r=0)=0$ and $P(r=R)=0,$ where $R$ is the radius of the surface of the star, one can compute the mass of the star via numerical integration of Eq.\eqref{massTOV} and Eq.\eqref{pressTOV}. This is shown in Fig.\eqref{Fig031}, for a QS made of MSQM, by assuming that the EoS can be extrapolated from the local volume $V$ to the whole star \cite{Martinez:2010sf}. Since the transveral pressure $P_{\perp}$ is the most sensible to the effects of the external magnetic field $\mathcal{B}$ and temperature $T$, $P_{\perp}$ was taken as the pressure in Eq.\eqref{pressTOV} to solve the system of equations. The MR relation was computed at a fixed  magnetic field $\mathcal{B}=5\times10^{18}$ G, $B_{\rm bag}=75$~MeV~fm$^{-3}$ and $T=0, 15, 30$~MeV (continuous, dashed and dotted lines). It is observed almost no change with temperature for the set of parameters chosen.

\section{Conclusions}

The thermodynamical properties of the MSQM in $\beta-$equilibrium, charge neutrality and baryon number conservation, have been presented in the previous section at zero and warm temperatures. The magnetic field effects are reflected in the particle energy spectrum, contributing mainly to the Landau diamagnetism of the magnetized gas of quarks and electrons. A possible nucleation phase transition may occur at strong manetic field values $\sim1.37\times10^{19}$ G, due to the sharp drop of the electron fraction, thus breaking the electric charge neutrality condition and possibly giving a new phase with strangelets. As the density increases at fixed values of $\mathcal{B}$ and $T$, the electron fraction becomes negligible and the quark fractions arrive to a same constant value indicating another possible phase transition with a negligible ammount of electrons and equal quark particle densities, like in the MCFL superconducting phase. The stability of the MSQM phase is enhanced by the external magnetic field at fixed densities, warm temperatures and Bag constant. Temperatures increase the energy per baryon as expected. The magnetic response of the MSQM shows the typical Landau diamagnetism behavior due to the quantization in Landau levels in the particle spectrum, with the corresponding increase of the magnetization with the temperature. This behavior enhances the pressure anisotropy of the MSQM, arriving to the equatorial collapse condition when $\mathcal{B}\sim1.37\times10^{19}$ G. The transversal pressure decreases with the temperature till arriving to a global density-dependent minimum temperature and beyond this point it starts to increase. However, this result has to be better studied including higher than the tree-level corrections to the thermodynamical potential. The EoS is as well presented showing positive and temperature-increasing speed-of-sound constants; the Bodmer-Witten-Terazawa hypothesis was shown to be true in this particular case. Finally, the Toldman-Oppenheimer-Volkof equations were solved with the aid of the obtained EoS to plot the Mass-Radius relation for stars made of MSQM, showing that warm temperatures don't make any significant change to this relation.

\end{document}